\documentclass[runningheads]{llncs}
\usepackage{hyperref}
\usepackage{graphicx}
\usepackage{xcolor}
\usepackage{bbm}
\usepackage{amsmath}
\begin{document}
\title{VesselShot: Few-shot learning for cerebral blood vessel segmentation}
%
%
\author{Mumu Aktar \inst{1}\orcidID{0000-0002-5718-7238} \and
Hassan Rivaz\inst{2}\orcidID{0000-0001-5800-3034}\and
Marta Kersten-Oertel\inst{1}\orcidID{0000-0002-9492-8402} \and
Yiming Xiao\inst{1}\orcidID{0000-0002-0962-3525}}
\authorrunning{M. Aktar et al.}
%
\institute{Department of Computer Science and Software Engineering, Concordia University 
\email{m\_ktar@encs.concordia.ca} \and
Department of Electrical and Computer Engineering, Concordia University
}
\maketitle              
\begin{abstract}
Angiography is widely used to detect, diagnose, and treat cerebrovascular diseases. While numerous techniques have been proposed to segment the vascular network from different imaging modalities, deep learning (DL) has emerged as a promising approach. However, existing DL methods often depend on proprietary datasets and extensive manual annotation. Moreover, the availability of pre-trained networks specifically for medical domains and 3D volumes is limited. To overcome these challenges, we propose a few-shot learning approach called VesselShot for cerebrovascular segmentation. VesselShot leverages knowledge from a few annotated support images and mitigates the scarcity of labeled data and the need for extensive annotation in cerebral blood vessel segmentation. We evaluated the performance of VesselShot using the publicly available TubeTK dataset for the segmentation task, achieving a mean Dice coefficient (DC) of 0.62$\pm$0.03.

\keywords{Few-shot learning, Deep learning, 3D volumes, Cerebrovascular segmentation}
\end{abstract}
\section{Introduction}
Cerebral blood vessel segmentation plays a vital role in various applications, such as diagnosing cerebrovascular diseases (e.g., stroke), planning surgical interventions for conditions, such as aneurysms and arteriovenous malformations, studying brain functions, and assessing the impact of new treatments for cerebrovascular disorders. However, publicly labeled data in this domain is limited, hindering the progress of deep learning-based research on cerebral blood vessel segmentation. Since 2017, numerous deep learning-based methods have been proposed for cerebral blood vessel segmentation. However, most of the previous work has been performed on data from private sources as there is very little publicly available labeled data~\cite{goni2022brain}. Traditionally, deep learning (DL) models for semantic segmentation require a large amount of training data with manual annotation, which is time-consuming and labor-intensive, particularly in 3D, where many slices must be inspected. The need for more annotated data negatively impacts DL models' training and generalization capabilities.

To address this challenge, few-shot learning emerges as a promising alternative that reduces the need for extensive manual annotation. For semantic image segmentation, few-shot learning aims to enable DL models to learn underlying visual patterns and semantics from a limited set of labeled examples. This also allows them to generalize effectively to unseen object categories during the segmentation process. To date, few-shot segmentation has been explored in several medical imaging contexts. In the study of Roy \emph{et al.}~\cite{roy2020squeeze}, the authors proposed a two-armed few-shot architecture to extract support and query images with squeeze-and-excitation modules for the segmentation of abdominal organs (each organ type is considered as a separate class) using 3D volumetric scans, obtaining an average Dice coefficient (DC) of 48.5\%~\cite{roy2020squeeze}. Similarly, Tang~\emph{et al.}~\cite{tang2021recurrent} used a few-shot framework to refine the segmentation masks with a recurrent module and achieved a mean DC of 81.91\%. To eliminate expert annotation for training medical image segmentation algorithms, Ouyang \emph{et al.}~\cite{ouyang2020self} employed a super-pixel-based self-supervised segmentation approach with few-shot learning. Preserving the local information to alleviate the foreground vs. background imbalance issue with an adaptive local prototype pooling, their study achieved a maximum DC of 78.84\%. Semi-supervised segmentation was also incorporated in a few-shot paradigm by considering a generative adversarial network (GAN)~\cite{mondal2018few}. This method had comparable performance to fully supervised approaches in multi-modal 3D medical image segmentation. Few-shot learning techniques have also shown efficacy in cardiac image sequence segmentation tasks following a multi-level semantic adaptation, with a DC of 92.43\%~\cite{guo2021multi}. Lastly, Xu \emph{et al.}~\cite{xu2022few} proposed a few-shot learning method with a multi-scale class prototype and attention module for 2D retinal blood vessel segmentation.

In this paper, we aimed to develop a few-shot learning approach, called ``VesselShot", for segmenting cerebral blood vessels.  Building upon the PANet few-shot segmentation method introduced by Wang \emph{et al.}~\cite{wang2019panet} for natural image segmentation based on metric learning, VesselShot leverages DL models' ability to learn a consistent embedding space that minimizes the distance between support and query prototypes (see Section ~\ref{problemdefinition}).  To the best of our knowledge, our method is the first attempt to employ few-shot learning for 3D segmentation of brain vascular images. The proposed VesselShot technique aims to overcome the limitations of the existing deep-learning models for cerebral blood vessel segmentation and explore the potential of few-shot learning in this domain.
\section{Methodology}
\subsection{Dataset and Pre-processing}
We used the publicly available TubeTK dataset \footnote{\url{https://public.kitware.com/Wiki/TubeTK/Data}}. Among the 100 magnetic resonance angiography (MRA) of healthy subjects, a subset of 42 have manual segmentation of the intracranial vasculature. The original dimension of the images is 448x448x128 voxels at a resolution of 0.5×0.5×0.8 \(mm^3\). The images were pre-processed as follows. First, all images were down-sampled to a resolution of 1x1x1 \(mm^3\), resulting in a dimension of 230x230x102 voxels. To allow spatial consistency, all the brains were registered to one subject's image as a template with affine transformations. Fifteen patches that contain blood vessels were randomly extracted from each brain using the technique introduced in the study of Wang \emph{et al.}~\cite{wang2020vc}, with a size of 64x64x16 voxels to fit the GPU memory. 
\subsection{Problem Definition}\label{problemdefinition}
To segment cerebral blood vessels with a small amount of annotated training data, we built upon the few-shot segmentation method proposed by Wang \emph{et al.}~\cite{wang2019panet}. In general, few-shot learning involves training and testing episodes with support and query sets, following a ``C-way K-shot" paradigm. The support set comprises labeled examples that a DL model can use to learn about target classes, while the query set contains unseen test cases to be classified during inference. In C-way K-shot segmentation, we obtain K \{image, mask\} pairs per semantic class in the support set, with a total of C classes. The training episodes consist of \(S_{i,k}\), \(M_{i, k}\) and \(Q_{i, k}\), denoting support, mask, and query sets, respectively with \(i=1,2,....c\) for \(c\) classes and \(k=1,2,...s\) for \(s\) samples/shots. Both the support and query sets share knowledge extracted by the DL model to perform the final segmentation. In our experiments, following the problem framing of Roy \emph{et al.}~\cite{roy2020squeeze}, who categorized classes with the designated segmentation tasks, we primarily focused on building our algorithm based on one class (i.e., blood vessel segmentation) or a 1-way K-shot approach. Furthermore, to account for individual vascular differences between subjects, we also considered the problem framing from the work of Xu \emph{et al.}~\cite{xu2022few}, who treated each subject as a separate class with its image patches as members of the class for retinal vessel segmentation. In this case, the segmentation was extended to a C-way K-shot setting.
\subsection{Model Design}
\begin{figure}[ht]
    \centering
    \includegraphics[width=\textwidth]{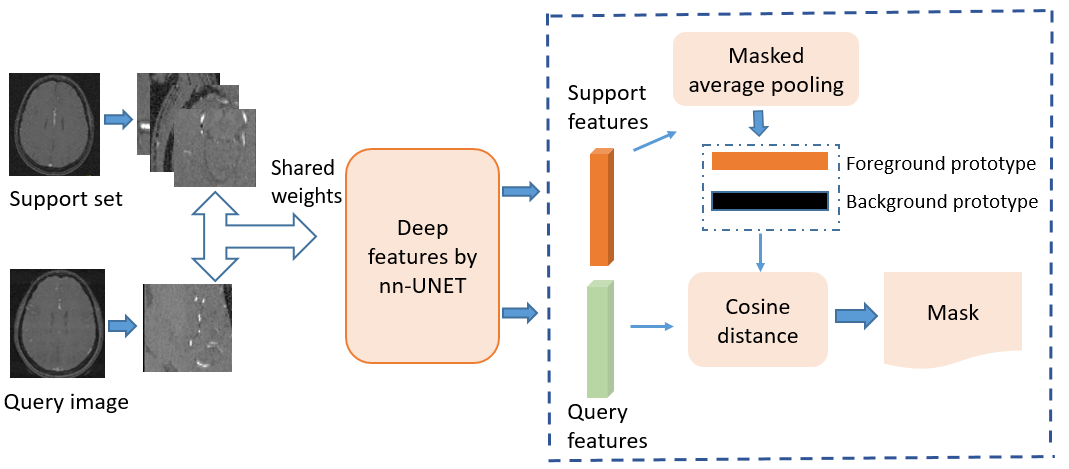}
    \caption{VesselShot 1-way 3-shot learning: 1 brain with 3 sample patches is considered for the support set. Knowledge is shared between support and query set by extracting deep features using nn-UNet~\cite{isensee2021nnu} which are further embedded into foreground and background prototypes using masked average pooling~\cite{wang2019panet}. Cosine similarity is used between support and query prototypes to obtain the segmented query mask.}
    \label{workflow}
\end{figure}

To perform support-to-query segmentation, we built robust prototypes from the target class of the support set. We used the nn-UNet~\cite{isensee2021nnu} architecture as a backbone network to extract deep features from support and query images. Upsampling along with masked average pooling~\cite{wang2019panet} was performed to obtain the final segmentation. Figure~\ref{workflow} shows an example of a 1-way 3-shot learning paradigm for query mask generation.

To generate a prototype from a class, the feature map, \(F_I\) of an image, \(I\) was extracted from the support set, \(S\) and by masked average pooling (Equation~\ref{eq:1}), the obtained feature maps were compared across different class indices of \(I\), similar to the prototype extraction approach used in PANet~\cite{wang2019panet}.

\begin{equation}\label{eq:1}
    prototype_{\text{class}} = \frac{1}{N_s} \sum_{i=1}^{N_s} \frac{\sum_{x,y,z}F_{I,\text{class}}^{(x,y,z)}\mathbbm{1}[Mask_{I,\text{class}}^{(x,y,z)} == \text{class}]} {\sum_{x,y,z} \mathbbm{1}[Mask_{I,\text{class}}^{(x,y,z)} == \text{class}]}
\end{equation}
where \(N_{s}\) is the total number of support images, and  $\mathbbm1$\([\cdot\)] takes a value of 1 when the condition  \([Mask_{I, class}^{(x,y,z)}==class]\) is true and 0 otherwise. The background prototype was built following the same equation with the constraint of \([Mask_{I, class} \neq class]\), which means the feature map values do not belong to the corresponding class index. For the evaluation, all brain class indices were assigned a value of 1 for foreground blood vessels (our main target) and 0 for the background. After prototype extraction, the feature map of the query image was compared with the support prototypes using cosine similarity. Each voxel at the spatial location (\(x,y,z\)) was classified based on the maximum similarity between the query feature map and the support prototypes. Finally, a segmentation mask was predicted for the query image based on the maximum probability values obtained by Softmax that was applied to the distance map. Note that our 3D segmentation was performed similarly to the work of Roy \emph{et al.}~\cite{roy2020squeeze}, but we experimented with the scenarios of 1-way K-shot and C-way K-shot as mentioned in Section~\ref{problemdefinition}. During inference time, we extracted 54 non-overlapping patches from a test MRA, where each was a query image at a given time and was paired with the support set to obtain the final segmentation. For our experiments, the support sets were created from the training data. 

For training, a combination of both cross-entropy loss, \(CE_{loss}\)  and Dice loss, \(D_{loss}\) was used. Our approach emphasized the Dice loss since blood vessels occupy a minimal area considering the brain space, which can significantly affect learning with a high-class imbalance. The Dice loss only focuses on the agreement of an image's predicted segmentation and ground truth label.  However, it is not ideal to overlook the background entirely, as this can affect the robustness of significant features of the network ~\cite{su2021dv}. Therefore, the following hybrid loss function was used to handle both class imbalance and increase the strength of features: $Loss =  0.6*CE_{loss}  + 0.7*D_{loss}$. Note that the weights were determined empirically.
\section{Experimental Setup} 
We employed the nn-UNet~\cite{isensee2021nnu} model to extract deep features from the support and query images. Since some fine-grained image features are lost while downsampling in feature extraction~\cite{xu2022few}, upsampling is an important step and pre-requisite for further background and foreground prototype extraction. Unlike Wang \emph{et al.}~\cite{wang2019panet}, our preliminary testing showed that nn-UNet~\cite{isensee2021nnu} was more effective as a decoder than trilinear upsampling for our 3D data (PANet used bilinear upsampling for 2D images~\cite{wang2019panet}).

To perform training in our few-shot blood vessel segmentation, a maximum iteration of 20,000 was used while monitoring the best DC value for early stopping. A learning rate (lr) scheduler with an SGD optimizer was used at an initial lr=0.01 and momentum of=0.99. For data augmentation, random flipping, random Gaussian blurring, noise addition, and contrast changes were applied during training. To evaluate the performance of our method, different metrics, including DC, precision, sensitivity, and Intersection over Union (IoU) were computed. We performed multiple experiments to prove the efficacy of the proposed few-shot segmentation method. These experiments encompassed various settings, namely 1-way 1-shot, 1-way 4-shot, 1-way 5-shot, and 3-way 5-shot learning. 
In addition, we also used a fully supervised UNet with four layers of hierarchies as a baseline to assess the performance of the proposed method using the same patch size and data augmentation techniques. It was trained using the Dice loss with an Adam optimizer $(lr=0.0001)$ and a CosineAnnealingLR scheduler $(T_{max}=5, eta_{min}=0.000001)$. To compare the different few-shot learning settings, as well as the UNet baseline,  we divided the data into three subsets: a training set (78\%, 33 cases), a validation set (7\%, 3 cases), and a test set (15\%, 6 cases). The best setting was determined from their performance based on the test set. Subsequently, we used the best setting to conduct a full 4-fold cross-validation. This way, we could obtain segmentation results for all the subjects to offer a more comprehensive evaluation.
\section{Results}
Table \ref{result_tab} presents the performance for various few-shot segmentation settings. It is important to note that the reported performance in the table was obtained from patch-based evaluations, where the averages of all classes in the test set were considered. The highest average DC of 0.67 was obtained with 1-way 1-shot learning. In terms of the mean values, 1-way 4-shot and 1-way 5-shot give similar results in all the evaluation criteria. However, the performance of the 3-way 5-shot method was notably inferior. This discrepancy may be attributed to the inclusion of three classes represented by distinct brains, characterized by significant similarities. Excess prototype generation in this approach likely contributed to overfitting, resulting in the observed decline in performance. The results of the 1-way 1-shot proved that the few-shot paradigm could offer sufficient segmentation performance with even a single sample from a single class, which makes faster convergence and mitigates the issue of a small annotated dataset. 

\begin{figure}[h]
    \centering
    \includegraphics[width=\textwidth]{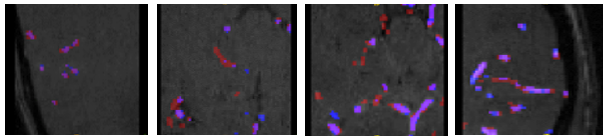}
    \caption{Segmentation maps of four samples, where red represents the original cerebral blood vessels, blue shows the prediction, and purple represents the overlap}
    \label{result}
\end{figure}

\begin{figure}
    \centering
    \includegraphics[width=\textwidth]{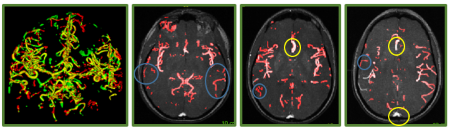}
    \caption{\emph{From left to right:} 3D segmentation result with the overlap of GT and predicted labels (yellow = overlap, green = GT and red = prediction). 2D Maximum intensity projections (MIPs) of 10 slices in Slice 15-25, 25-35, and 35-45 from a total of 102 brain slices, with the overlay of the original MRA and segmentation (in red). The blue circles show the wrong prediction of large vessels and the yellow circles indicate missed blood vessels.}
    \label{3d}
\end{figure}

It is essential to note that while the single-split result indicated a slight advantage for the 1-way 1-shot model in the case of DC, the more comprehensive evaluation through cross-validation provided a more precise and more reliable picture of the model's performance. Therefore, considering all performance metrics, the 1-way 5-shot setting emerged as the top-performing setting among the options tested in this study. Based on the 1-way 5-shot setting, a full 4-fold cross-validation was performed. The metrics of DC, sensitivity, precision, and IoU were obtained as 0.62$\pm$0.03, 0.53$\pm$0.02, 0.72$\pm$0.02, and 0.43$\pm$0.02, respectively. For qualitative evaluation, segmentation maps of four random patches are shown in Fig.~\ref{result} for the 1-way 5-shot setting. Furthermore, by recombining segmented image patches from the same brain spatially, we also demonstrate a case in Fig.~\ref{3d} for the same setting. Finally, the fully supervised UNet performed poorly with the limited annotated data and achieved a DC of 0.27$\pm$0.27. UNet typically requires a larger well-labeled dataset to achieve reasonable performance, as demonstrated by the study of Livne \emph{et al.}~\cite{livne2019u}.
\begin{table}[h]
     \caption{Performance metrics of VesselShot for different settings with the UNet as a baseline, including DC, Sensitivity, Precision, and IoU.}\label{result_tab}
    \centering
    \begin{tabular}{|p{2.5cm}|p{2cm}|p{2.5cm}|p{2.5cm}|p{2.2cm}|} \hline
       \textbf{Methods} & \textbf{DC (SD)} & \textbf{Sensitivity(SD)} & \textbf{Precision(SD)} & \textbf{IoU(SD)} \\ \hline
        1-way 1-shot& \textbf{0.67(0.02)}  & 0.50 (0.03) & 0.68 (0.02) & 0.40 (0.02)  \\ \hline
        1-way 4-shot& 0.66 (0.02) & 0.54 (0.03) & \textbf{0.68 (0.02)} & 0.41 (0.02)\\ \hline
        1-way 5-shot& 0.66 (0.02) & 0.58 (0.02) & \textbf{0.71 (0.03)} & \textbf{0.45 (0.02)} \\ \hline
        3-way 5-shot& 0.52 (0.04) & 0.39 (0.04) & 0.57 (0.08) & 0.23 (0.03) \\ \hline
        UNet & 0.27 (0.27) & 0.47 (0.09) & 0.35 (0.06) & 0.15 (0.04) \\ \hline
    \end{tabular}
\end{table}
\section{Discussion}
This paper proposed a novel few-shot learning approach for 3D cerebral blood vessel segmentation. The method achieves the segmentation by building robust prototypes with masked average pooling based on embedded features that are extracted from an nn-UNet~\cite{isensee2021nnu}. Inspired by the PANet~\cite{wang2019panet}, we adopted foreground and background prototypes and used them to compare the query feature map with the support set's prototypes for blood vessel segmentation. To further enhance generalizability, Wang et al.~\cite{wang2019panet} performed prototype alignment regularization (PAR) of the predicted query mask with the support mask through additional information extraction. We have also experimented with this technique for our application, but unfortunately, it did not lead to performance gains during evaluation and resulted in slower convergence in model training.

In our approach, we considered two scenarios: (1) a single class (blood vessel segmentation) in a 1-way K-shot setting and (2) a C-way K-shot setting with each brain as a separate class. While the first case aligns with the approach of Roy \emph{et al.}~\cite{roy2020squeeze}, the latter resembles the problem framing of Xu \emph{et al.}~\cite{xu2022few}. In the 1-way K-shot setting, the best results came from the 1-way 5-shot setting, which was superior to treating different brains as their own classes. We hypothesize that this was due to the high structural similarity between MRAs after spatial normalization, which emphasizes the primary vasculature networks. This is in contrast to the results of Xu \emph{et al.} ~\cite{xu2022few}. Since the core task involves only two classes, blood vessels, and the background, the C-way K-shot paradigm may lead to overfitting and compromise performance. In the future, we will continue to explore different framings of the C-way K-shot setup for improved accuracy. For example, treating 3D image patches from consistent spatial locations in a stereotactic space as distinct classes to allow enhanced feature encoding.

We found that misclassifications predominantly occurred near the brain surface, where surface veins and the dura (both with bright signals) reside. This was partially due to the fact that the manual ground truths of the MRA segmentation primarily focus on the main arteries rather than the surface vasculatures, which is of interest in neurosurgical planning~\cite{hellum2022novel,beriault2012towards}. The misclassifications may also be caused by training the model with random patches that were mostly taken from the center of the brain. In the future, we will incorporate random patches that consider both vessel and non-vessel regions, along with an increased number of patches.

The recent work by Li \emph{et al.}~\cite{li2022gvc} introduced a global vascular context network (GVC-Net) with a hybrid loss to address over-segmentation issues caused by sparse labels and skull vessels. They also utilized the TubeTk dataset by training on 42 data points and testing on 10 data points, achieving a sensitivity of 61.24\%, precision of 75.58\%, DC of 67.66\%, intersection over the union of 51.13\%, and centerline Dice coefficient of 83.79\%. Although their method has a 5\% higher Dice coefficient, it is important to note that their reported result is based on a single fold while our method's performance was from a full 4-fold cross-validation. In a separate study, Tang \emph{et al.} \cite{tang2021recurrent} compared their proposed RPNet approach to PANet \cite{wang2019panet} in 3D abdominal image segmentation. RPNet outperformed PANet, achieving approximately 33\% higher performance. Given these promising results, it would be worthwhile to investigate RPNet's application for 3D cerebral blood vessel segmentation task.

One major benefit of few-shot learning is the capacity to allow high flexibility and adaptability for unseen classes, which can include new classification/segmentation tasks and image contrasts. Our proposed approach shows sufficient generalizability to new classes as only a few image patches in the support set allowed the segmentation of the whole brain volume. In contrast, Holroyd \emph{et al.}~\cite{holroyd2023tube} developed tUbe net, a model that achieved high performance in segmenting new blood vessels through transfer learning. However, tUbe net requires a large training dataset, which may not always be available. Our method is independent of pre-trained weights and can be potentially applicable to diverse applications.

Despite efforts to utilize limited annotated data, the current performance of VesselShot is not sufficiently accurate for clinical deployment. However, we will explore the strategies mentioned above to improve the accuracy and robustness of few-shot cerebral vascular segmentation. Despite its limitations, our proposed method represents a preliminary step in addressing limited annotated data in the challenging task of 3D cerebral blood vessel segmentation.
\section{Conclusion}
Our novel method utilizes few-shot learning to address the challenges of limited labeled datasets in 3D cerebral blood vessel segmentation. This approach shows promise in overcoming the bottleneck of limited manually annotated datasets and could aid in clinical tasks with further improvement in the future. While the present achievement may not yet find direct application in clinical environments, it signifies an advancement in this domain.
\section*{Acknowledgements}
This study was funded by an FRQNT Team Grant (2022-PR-296459).

\bibliographystyle{IEEEtran}
\bibliography{samplepaper_1.bbl}
\end{document}